\documentclass[aps,prd,preprint,showpacs]{revtex4}
\usepackage{graphicx}
\begin{document}
\draft
\title{Effects of the CP Odd Dipole Operators on Gluino-Squark Production }
\author{Ahmet T. Alan}
\thanks{e-mail: alan\_a@ibu.edu.tr}
\affiliation{Department of Physics, Abant Izzet Baysal University,
14280, Bolu, Turkey}

\begin{abstract}
We analyze effects of the CP violating interactions on associated
gluino-squark production in the MSSM at hadron colliders. Depending
on the sparticle masses, the hadronic cross sections can be enhanced
up to 9.5 \% for a 500 GeV gluino at the LHC energies.

\end{abstract}
\maketitle Standard Model (SM) has the only source, CKM phase for CP
violation and is unable to predict the baryon asymmetry of the
universe. Supersymmetric models with minimal field content (MSSM),
on the other hand, allow for presence of several CP violating phases
\cite{Brhlik:1998zn, Barger:2001nu, Demir:2003js}. These models are
constrained by the experimental bounds on the Electric Dipole
Moments (EDMs) \cite{Bartl:1999bc, Abel:2001vy, Huber:2006ri}. Thus,
EDMs are prime observables for testing new physics beyond the SM and
the CP violating interactions inducing them must be included in the
analyzes of reactions involving colored sparticles at, in
particular, the CERN Large Hadron Collider (LHC). In a previous
work, we have analyzed effects of these interactions in the pair
production of gluinos at hadron colliders \cite{Alan:2005gu}.
Another potentially important gluino production process, related to
that work, is the gluino-squark associated production. It has been
shown in \cite{Alan:2007rp} that gluino-squark pair production is
independent of the SUSY CP-odd phases at tree level. Therefore,
CP-odd phases show up in production cross sections only at the loop
level. Consequently, inclusion of the CP-odd dipole operators in
scattering amplitudes provides an effective parametrization of the
CP-odd phase effects. Our aim in this brief report is to analyze
effects of these operators, which induce the color electric dipole
moments (CEDMs) of the quarks and gluinos, on the hadronic
production cross sections at the Tevatron and LHC with center of
mass energies of 1.8 and 14 TeV, respectively.

\begin{figure}[h]
\centering
  \includegraphics[width=12cm]{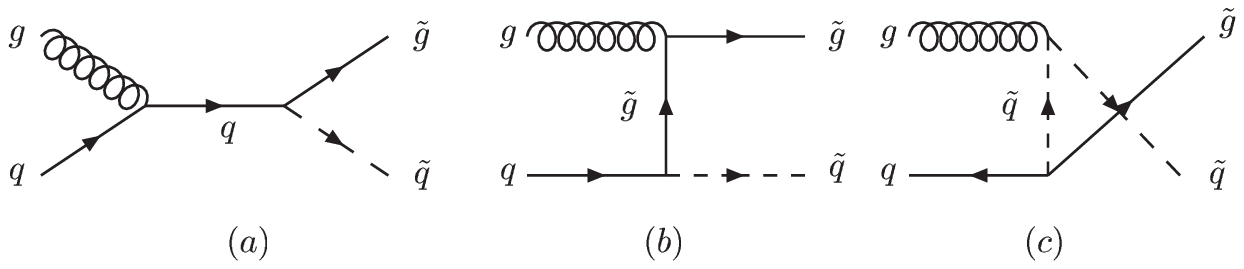}\\
 \caption{The Feynman diagrams for gluino-squark production in $qg$ scattering }
 \label{dia}
\end{figure}
In $pp$ ( or $p\bar p$) collisions, the gluino-squark production
occurs via the subprocess $qg\rightarrow \tilde g\tilde q$ through
the three s,t and u channels as shown in Fig.~\ref{dia}. Including
the CP violating effective dimension five operators which generate
the CEDMs of quarks and gluinos the following SUSY-QCD interaction
vertices arise in the calculation of the parton level differential
cross section :\\
i) Quark-quark-gluon\\
 \includegraphics[width=9cm]{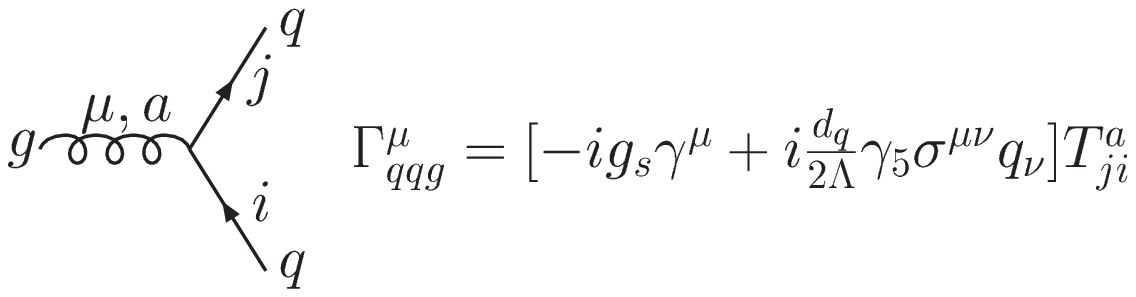}\\
ii)Gluon-gluino-gluino\\
\includegraphics[width=9cm]{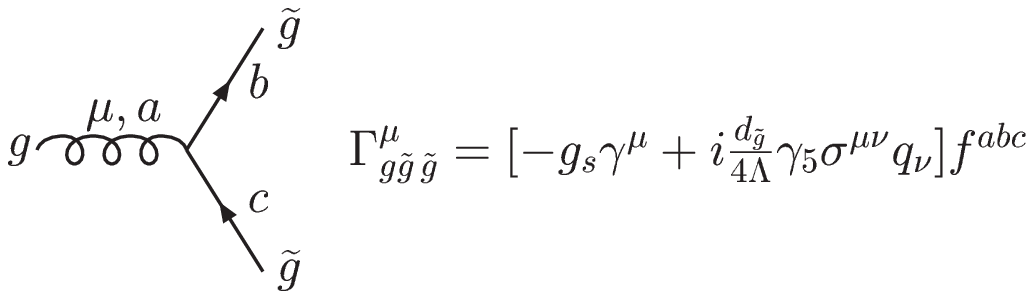}\\
iii)Quark-squark-gluino\\
\includegraphics[width=9cm]{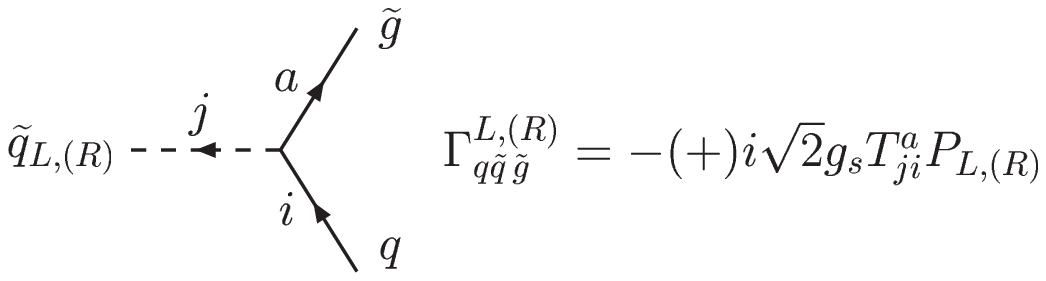}\\
iv) Gluon-squark-squark\\
\includegraphics[width=9cm]{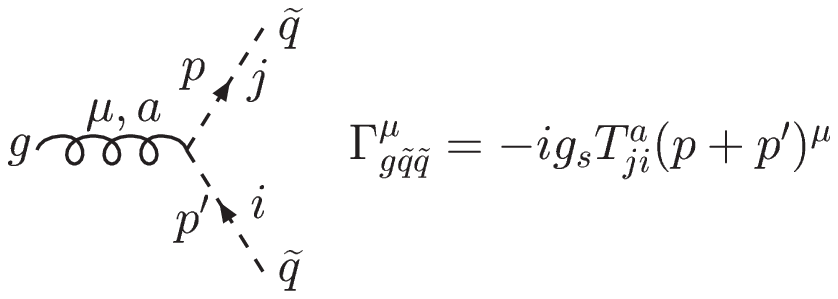}\\
where $q_{\nu}$ is the gluon momentum, $d_{q}$ and $d_{\tilde g}$
are the CEDMs of quarks and gluinos, respectively, $T^a$ are the
generators of $SU(3)_c$ color symmetry group, $f^{abc}$ are
structure constants and $\Lambda$ is a scale up to which effective
theory is valid.

The color and spin averaged parton level cross section for $\tilde g
\tilde q$ production is given by
\begin{eqnarray}\label{csf}
 \hat{\sigma}_q (qg\rightarrow \tilde g\tilde q)&=& \frac{\pi\alpha_s^2}{9\hat s^3}\Bigg[
2(m_{\tilde g}^2-m_{\tilde q}^2)(4m_{\tilde g}^2+5m_{\tilde
q}^2-4\hat s)\ln[\frac{m_{\tilde g}^2-m_{\tilde q}^2-\hat
s+J}{m_{\tilde g}^2-m_{\tilde q}^2-\hat s-J}]\nonumber\\ &+& 9(\hat
s^2+2(m_{\tilde q}^2-\hat s)(m_{\tilde q}^2-m_{\tilde
g}^2))\ln[\frac{m_{\tilde g}^2-m_{\tilde q}^2+\hat s+J}{m_{\tilde
g}^2-m_{\tilde q}^2+\hat s-J}]\nonumber\\&+&J(32(m_{\tilde
q}^2-m_{\tilde g}^2)-7\hat s )\Bigg]
+\frac{\alpha_sJ}{8\Lambda^2\hat s^2}\left(\frac{d_{
q}^2}{9}+\frac{d_{\tilde g}^2}{16}\right)(\hat s+m_{\tilde
g}^2-m_{\tilde q}^2)
\end{eqnarray}
where $\hat s=\tau S$ is the partonic center of mass energy and J is
given by \[J=\sqrt{(m_{\tilde g}^2-m_{\tilde q}^2)^2-2\hat
s(m_{\tilde g}^2+m_{\tilde q}^2)+\hat s^2}.\] The hadronic cross
section $pp (p\bar p)\rightarrow \tilde g \tilde q X$ is calculated
as a function of hadronic center of mass energy $S$, superpartner
masses $m_{\tilde g}$ and $m_{\tilde q}$, by convoluting the
subprocess cross section:
\begin{eqnarray}
\sigma_{tot} (pp (p\bar p)\rightarrow \tilde g \tilde q
+X)&=&\int_{\frac{(m_{\tilde g}+m_{\tilde q})^2}{S}}^1d\tau
\int_\tau^1 \frac{dx}{x}\frac{1}{1+\delta_{ij}}\sum_{i j}[f_i
(x,Q^2)f_j(\frac{\tau}{x},Q^2)\nonumber\\
&+&f_i(\frac{\tau}{x},Q^2)f_j(x,Q^2)]\hat\sigma(qg\rightarrow \tilde
g\tilde q)
\end{eqnarray}
where $(i,j)=(g, q)$ the structure function $f_i(x,Q^2)$ represent
the number density of parton $i$ carrying the fraction $x$ of the
longitudinal proton momentum. In calculation of the hadronic cross
sections we used the CTEQ \cite{Lai:1994bb} parametrization of the
five light quarks and gluons in proton. All couplings and masses in
the partonic reactions are defined at the factorization scale Q,
which has to lie around $m_{\tilde g}+m_{\tilde q}$.

\begin{figure}
  \includegraphics[width=12cm]{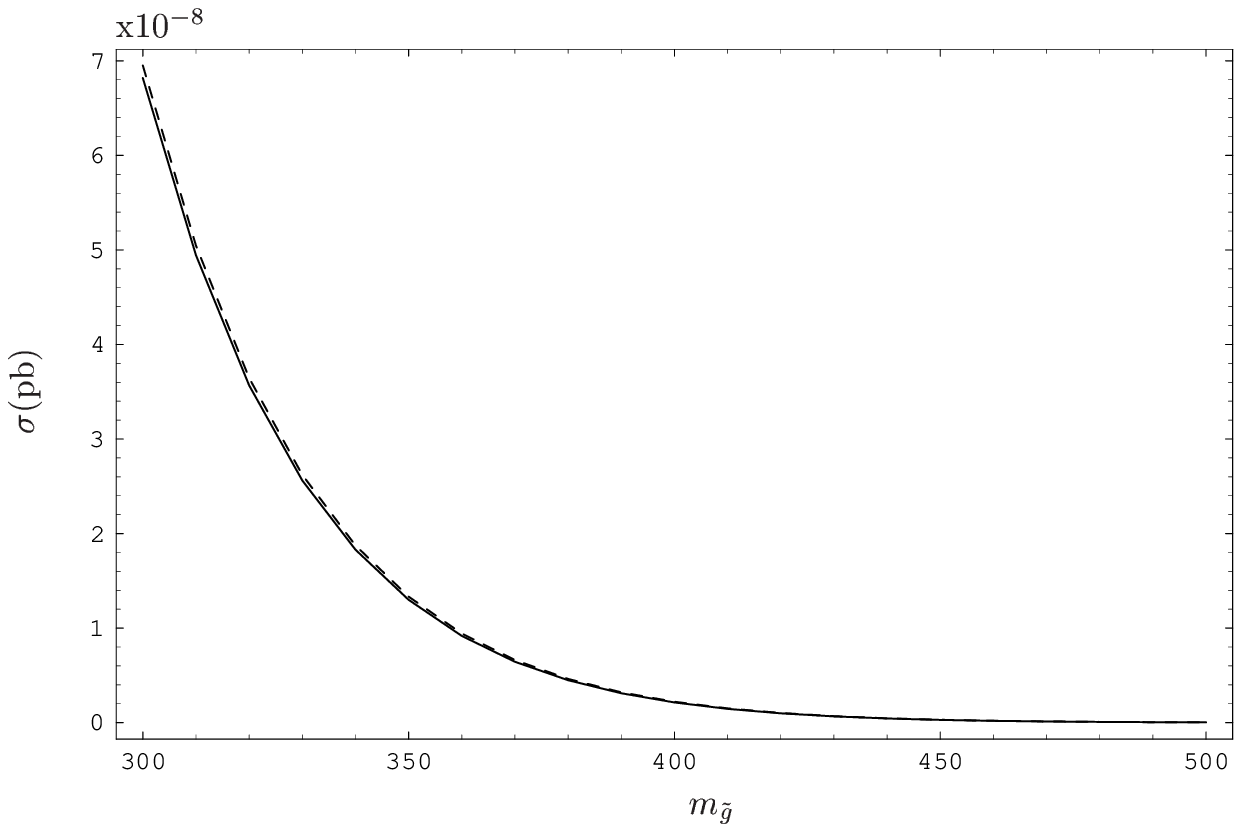}\\
  \caption{The total cross sections for $p\bar
p\rightarrow \tilde g \tilde q X$ at the Tevatron for
$\Lambda$=$m_{\tilde q}$=1 TeV.}\label{fig1}
\end{figure}

\begin{figure}
  \includegraphics[width=12cm]{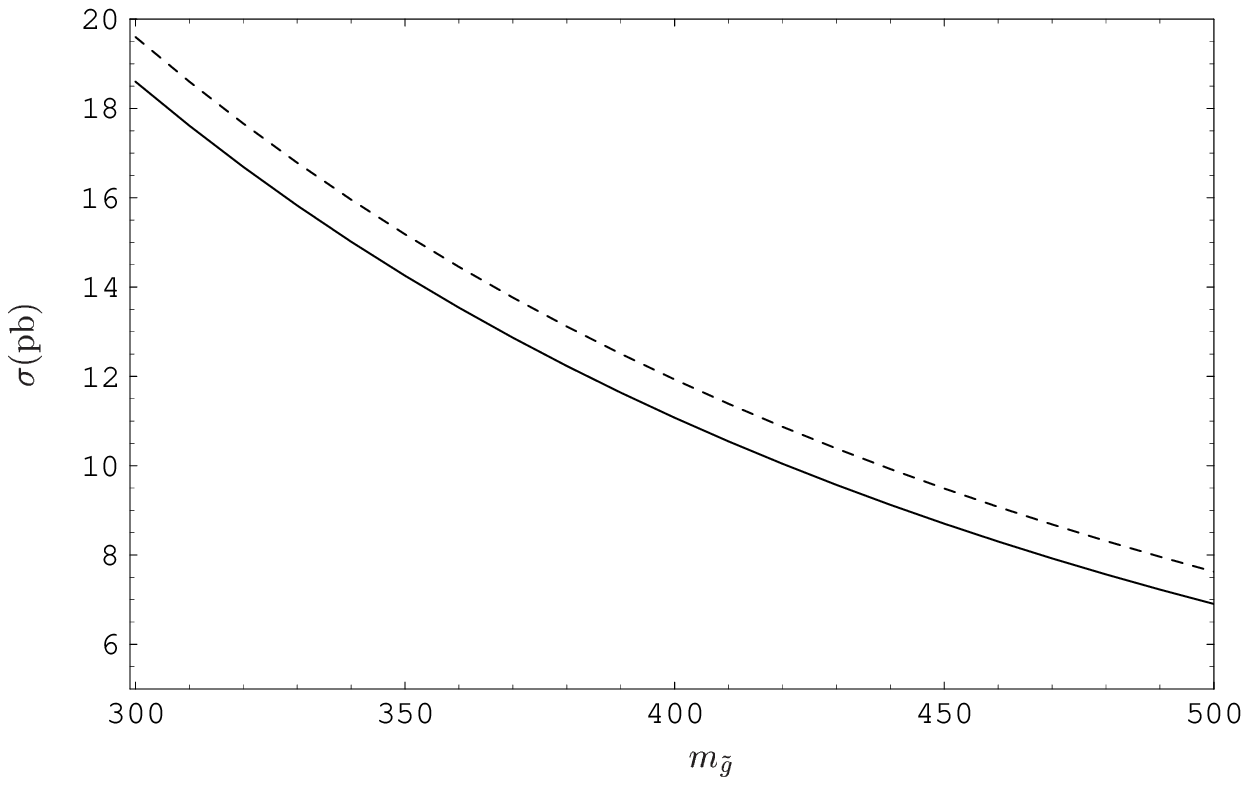}\\
  \caption{The total cross sections for
$pp\rightarrow \tilde g \tilde q X$ at the LHC for
$\Lambda$=$m_{\tilde q}$=1 TeV.}\label{fig2}
\end{figure}

In Tables \ref{tab1} and \ref{tab2} we have presented the cross
sections for the gluino masses of 300, 400 and 500 GeV by taking
$m_{\tilde q}$= 1 TeV. The values in the last rows of the tables
were obtained for $\Lambda$=$m_{\tilde q}$=1 TeV and $d_{\tilde
g}$=$d_{q}$=0 and were presented to displayed the effects of CP-odd
operators explicitly by comparing them with the ones in the second
rows of the tables. The results for the $pp (p\bar p)\rightarrow
\tilde g \tilde q X$ cross sections are presented in
Figs.~\ref{fig1} and \ref{fig2} at the Tevatron and LHC,
respectively again by fixing $\Lambda$=$m_{\tilde q}$=1 TeV. The
solid lines represent the $d_{\tilde g}$=$d_{ q}$=0 case and dotted
lines correspond to added the contributions of CP-odd operators. For
illustrations the CEDMs $d_{\tilde g}$ and $d_{ q}$ are set equal to
1. Since Eq.~(\ref{csf}) is even in $d_{\tilde g}$ and $d_{ q}$ the
choice of sign for these coefficients is immaterial for
$qg\rightarrow \tilde g\tilde q$ cross sections.

The enhancements in the hadronic cross sections at Tevatron for
gluinos of masses 300, 400 and 500 GeV are  1.94 \%, 3.84 \% and
5.19 \%, respectively. The event rates at this collider is so slow
that it yields 1-2 event per year for 500 GeV gluinos.

At the LHC, the event rates are very high, for instance for 500 GeV
gluinos, the event number is $10^7$ in each LHC detector with an
integrated luminosity of 100 $fb^{-1}$. To these high event rates,
CP violating interactions provide extra contributions yielding the
enhancements of 5.38 \%, 7.73 \% and 9.48 \% in the hadronic cross
sections, for 300, 400 and 500 GeV gluinos, respectively.

\begin{acknowledgements}
This work is partially supported by Abant Izzet Baysal University
Research Fund. I am grateful to D. A. Demir for useful discussions.
\end{acknowledgements}

\begin{table}[h]
  \centering
\begin{tabular}{c|ccc}
  \hline\hline
  $\Lambda$ GeV &$m_{\tilde g}$=300 GeV &$m_{\tilde g}$=400 GeV  & $m_{\tilde g}$=500 GeV  \\
  \hline
  800& 7.0251x10$^{-8}$&2.2484x10$^{-9}$&3.3199x10$^{-11}$ \\
  1000&6.9506x10$^{-8}$&2.2082x10$^{-9}$&3.2214x10$^{-11}$\\
  2000&6.8513x10$^{-8}$&2.1546x10$^{-9}$&3.1021x10$^{-11}$ \\
  4000&6.8265x10$^{-8}$&2.1412x10$^{-9}$&3.0723x10$^{-11}$ \\
  6000&6.8219x10$^{-8}$&2.1387x10$^{-9}$&3.0668x10$^{-11}$ \\
  8000&6.8203x10$^{-8}$&2.1379x10$^{-9}$&3.0649x10$^{-11}$\\
\hline\hline $d_{\tilde g}=d_{
q}=0$&6.8183x10$^{-8}$&2.1367x10$^{-9}$&3.0624x10$^{-11}$\\
\hline\hline
\end{tabular}
\caption{The hadronic cross sections in pb at the Tevatron ($\sqrt
S$= 1.8 TeV) for $m_{\tilde q}$=1 TeV}\label{tab1}
\end{table}
\begin{table}[h]
\centering
\begin{tabular}{c|ccc}
  \hline\hline
  $\Lambda$ GeV & $m_{\tilde g}$=300 GeV & $m_{\tilde g}$=400 GeV  &$m_{\tilde g}$=500 GeV  \\
  \hline
  800& 20.1639&12.4119&8.0331\\
  1000&19.6009&11.9303&7.6266\\
  2000&18.8503&11.2882&7.0846 \\
  4000&18.6627&11.1277&6.9491\\
  6000&18.6279&11.0979&6.9240\\
  8000&18.6158&11.0875&6.9152 \\
\hline\hline
 $d_{\tilde g}=d_{
q}=0$&18.6001&11.0742&6.9039 \\
\hline\hline
\end{tabular}
\caption{The hadronic cross sections in pb at the LHC ($\sqrt S$= 14
TeV) for $m_{\tilde q}$=1 TeV.}\label{tab2}
\end{table}
\end{document}